\documentstyle[multicol,aps,prl,eqsecnum,epsf]{revtex}
\begin{document}
\title{The Scalings of Scalar Structure Functions in a Velocity Field
with Coherent Vortical Structures}
\author{M A I Khan and  J C Vassilicos}
\address{\em Department of Applied Mathematics and
Theoretical Physics, Silver Street, University of Cambridge,\\
Cambridge CB3 9EW, United Kingdom}
\date{March 17, 2000 Revised 13 September 2001}
\maketitle
\begin{abstract}
In planar turbulence modelled as an
isotropic and homogeneous collection of 2-D non-interacting compact
vortices, the structure functions ${\cal S}_{p}(r)$ of a statistically
stationary passive scalar field have the following scaling behaviour
in the limit where the P\'eclet number $Pe\rightarrow\infty$
\vspace{-2mm}
\[{\cal S}_{p}(r)\sim constant+\ln\left({\frac{r}{LPe^{-1/3}}}\right)\ 
{\rm{for}}\ LPe^{-1/3}\ll r\ll L\ , \]
\vspace{-5mm}
\[{\cal S}_{p}(r)\sim  \left({\frac{r}{LPe^{-1/3}}}\right)^{6(1-D)}\
{\rm{ for}}\ LPe^{-1/2}\ll r\ll LPe^{-1/3}, \]
\noindent where $L$ is a large scale and $D$ is the fractal co-dimension
of the spiral scalar structures generated by the
vortices ($1/2\leq D<2/3$). Note that $LPe^{-1/2}$ is the scalar Taylor
microscale which stems naturally from our analytical treatment of the
advection-diffusion equation. The essential ingredients of our theory
are the locality of inter-scale transfer and Lundgren's time average
assumption. A phenomenological theory explicitly based only on these two
ingredients reproduces our results and a generalisation of this
phenomenology to spatially smooth chaotic flows yields $(k\ln k)^{-1}$
generalised power spectra for the advected scalar fields.  
\end{abstract}
\begin{multicols}{2}
\section{Introduction}
The theory of turbulent passive scalars has received much attention
recently\cite{Kr94,Kup95,Cher96}. The mixing of a scalar
field $\theta$ in a velocity field ${\bf v}$ is governed by the
advection-diffusion equation
\begin{equation}
\partial_{t}\theta({\bf x},t)+{\bf v}({\bf x},t)\cdot{\bf \nabla}
\theta({\bf x},t)=\kappa \nabla^{2} \theta({\bf x},t)+f({\bf x},t)
\label{diffeqn1}
\end{equation}
where $\kappa$ denotes the molecular diffusivity of the scalar
$\theta$ and $f({\bf x},t)$ is an external source ( forcing) driving the
scalar. The mixing of a scalar field $\theta$ is characterised by its
structure functions \(S_{p}\equiv\langle[\theta({\bf {x}}+{\bf{r}})-\theta({\bf{ x}})]^{p}\rangle=\langle(\delta\theta({\bf
r}))^{p}\rangle\) for any number $p$. If we want to find $S_{p}$ then
we need models of the velocity field ${\bf v}$.  
 
The model which has attracted much attention recently is the
Kraichnan model\cite{Kr94} where the velocity field ${\bf v}$ is
considered to be incompressible, statistically isotropic, white-noise
in time($\delta$-correlated) and Gaussian. Furthermore it has
homogeneous increments with power law spatial correlations
\begin{eqnarray}
&&\langle[v_{i}({\bf r},t)-v_{i}(0,0)][v_{j}({\bf
r},t)-v_{j}(0,0)]\rangle=\nonumber\\&&
2\delta(t)r^{h}\left[(h+d-1)\delta_{ij}-h{\frac{r_{i}r_{j}}{r^{2}}}\right]\nonumber\\
\label{velincr}
\end{eqnarray}
where the scaling exponent $h\in]0,2[$ and $d$ is the dimension of
space so that $i,j=1,2,\ldots,d$. The above tensorial structure of the
velocity field is in conformity with incompressibility. The Kraichnan
model also assumes a forcing $f({\bf x},t)$ in (\ref{diffeqn1}) that
is an independent Gaussian random field with zero mean. The forcing is
white in time and its covariance is assumed to be a real, smooth,
positive-definite function with rapid decay in space so that the
forcing is homogeneous, isotropic and takes place on the integral
scale $L$. The generic scaling behaviour of the structure function
\(S_{p}\sim r^{\zeta_{p}}\), \(r\ll L\), of passive scalars in the
Kraichnan model was established in\cite{Kr94,Kup95,Cher96,Boris}. The scaling exponents
of this formalism are of the form \(\zeta_{p}=\zeta_{p}(d,h,p)\) where
\(h \in [0,2]\) is the H$\ddot{o}$lder exponent in (\ref{velincr}). In
the context of this  model Balkovsky and Lebedev \cite{Bal98} and
Chertkov\cite{Cher97} used the instantonic formalism in a $d$-
dimensional space to find the scaling exponents for large $p$.  It was
also shown in the instantonic formalism that
\(\lim_{p\rightarrow\infty}\zeta_{p}\simeq
\frac{d(2-h)^{2}}{8h}\)\cite{Bal98} which is independent of $p$. The
scaling exponents were also calculated using other techniques in the
limits $h\rightarrow 0$\cite{Kup95,96BGK},
$d\rightarrow\infty$\cite{CFKL95,Cher96} and
$p\rightarrow\infty$\cite{Bal98}, and a $2-h$ expansion of $\zeta_{p}$
was proposed in \cite{Boris}. It was found that $\zeta_{p}$ does
depend on $p$ for small values of $p$ in the Kraichnan model.

Our work lies in the opposite extreme of the Kraichnan model. We work
in the regime where we have a persistent vortical velocity field
frozen in time in two dimensions. The important differences between
this model and the Kraichnan model are in the structure of the
velocity field infinitely correlated in time in this model but delta
correlated in time in the Kraichnan model; and vortical in space in
this model, but Gaussian in the Kraichnan model. In this model, the
velocity field is that of spatially distributed noninteracting two
dimensional vortices with compact structure. We consider the spatial
distribution of vortices to be dilute in that they are far from each
other and therefore maintain their structure and spatial position for
an indefinite period. We also consider this distribution to be
homogeneous and isotropic and the velocity field to be incompressible,
that is $\nabla\cdot{\bf v}=0$. The model of the velocity field
considered here is an artificial model of planar homogeneous
turbulence where the emphasis is on the coherent vortex aspect of the
flow. In order of presentation, the first aim of this model is to
demonstrate that in the case of the unforced scalar ($f=0$ in
\ref{diffeqn1}) we can quantify the statistics of the turbulent scalar
field in terms of the scalar's spiral geometry generated by the
coherent vortical structures in the flow(sections \ref{secsf} and
\ref{secmulsp}). The second aim is to derive the Batchelor $k^{-1}$
power spectrum and all the corresponding structure functions for the
scalar field in the case where the scalar is forced and in a
statistically steady state (section \ref{avltm}). Such a spectrum has
been recently observed by Jullien et. al.,\cite{JCT99} in a 2-D
turbulent flow with well defined, albeit short lived, coherent
vortical structures.

In the next section we discuss the scenario of a decaying scalar field
in an isolated vortex. In section \ref{secsf}, we define structure
functions and calculate the spectrum of higher order correlation
functions for a single spiral created by a single vortex and decaying
by the action of molecular diffusion. In section \ref{secmulsp} we
generalise our analysis to many non-interacting vortices and calculate
the scalings of the structure functions of the decaying scalar
field. In section \ref{avltm} we calculate the generalised power
spectra and the corresponding structure functions of statistically
stationary  scalar spirals by applying the time-average operation
approach of Lundgren\cite{Lund82}. In section \ref{pheno} we discuss
the phenomenology behind the $k^{-1}$ scalings of the generalised
power spectra. Section \ref{disc} contains conclusions, discussion and
the obtainment of the $(k\ln k)^{-1}$ scalings of the generalised
power spectra in smooth chaotic flows. 


\section{Passive scalar in a  planar vortex}\label{sec1vor}
The advection of a decaying passive scalar field by a single planar
vortex has been studied by Flohr and Vassilicos\cite{Flohr}. We use the
formulation used in \cite{Flohr} namely:
\begin{equation}
\partial_{t}\theta + \Omega(r)\partial_{\phi}\theta= \kappa\nabla^{2}\theta
\label{diffeqn3}
\end{equation}
where \(\partial_t= \displaystyle{\frac{\partial}{\partial t}}\) and
\(\partial_{\phi}= \displaystyle{\frac{\partial}{\partial \phi}}\) and
\(\Omega(r)=\Omega_{0}\left(\frac{r}{L}\right)^{-s}\) and $L$ is the
maximum distance of the scalar interface from the centre of the
vortex. This equation describes the advection and diffusion of a
scalar field $\theta$ in the azimuthal plane \({\bf r}=(r, \phi)\) by
a steady vortex with  azimuthal velocity component
\(u_{\phi}(r)=r\Omega(r)=L\Omega_{0}\left(\frac{r}{L}\right)^{1-s}\)
and vanishing radial and axial velocity components. Direct Numerical
Simulations and experiments in the laboratory have demonstrated the
existence and importance of coherent vortices in two-dimensional
turbulence and in two-component turbulence in stably-stratified flow
with and without rotation of the reference frame
\cite{Williams,Ambaum,Holton,WMGolub}. Note that  axial velocity
fields of the form \(u_{\phi}(r)=
L\Omega_{0}\left(\frac{r}{L}\right)^{1-s}\) have been used in
\cite{Lund82,Flohr,Gilbert,Segel,98PuSaf,97VNIK,99AV,PV2000} and that
their large wavenumber energy spectrum has the form $E(k)\sim
k^{-5+2s}$ for $1/2<s<2$ with the appropriate large scale bound.  We
choose $s\geq 1$ to ensure that $u_{\phi}(r)$ does not increase with
increasing $r$ and $s<2$ to ensure that the energy spectrum is steeper
than $k^{-1}$. The initial scalar field \(\theta_{0}=\theta({\bf
x},t=0)\) is characterised by a regular interface between
$\theta_{0}\neq 0$ and $\theta_{0}=0$ with minimal distance $r_{0}$
and maximal $L$ from the rotation axis. By regular structure we mean
that the interface has no irregularities on scales smaller than
$L$. Nothing else needs to be specified about the initial scalar field
$\theta_{0}({\bf x})$. Such a patchy initial condition where all the
non-zero scalar is confined within a regular interface mimics well
initial conditions in certain laboratory experiments where scalar is
injected in the flow in the form of blobs(e.g. \cite{JCT99}).

As time proceeds, the patch winds around the vortex and builds up
a spiral structure and decays due to diffusion. The characteristic
time $\Omega_{0}^{-1}$ is the inverse angular velocity of the vortex
at $L$. This defines a P\'eclet number
\(Pe=\Omega_{0}L^{2}\kappa^{-1}\). We non-dimensionalise equation
(\ref{diffeqn3}) by using the following transformations
\[L^{-1}r\rightarrow r, \Omega_{0}t\rightarrow t,
\Omega_{0}^{-1}\Omega(r)\rightarrow \Omega(r),
L^{2}\nabla^{2}\rightarrow \nabla^{2},\] and equation (\ref{diffeqn3})
takes the form
\begin{equation}
\partial_{t}\theta + \Omega(r)\partial_{\phi}\theta=
{\frac{1}{Pe}}\nabla^{2}\theta\ .
\label{ndeqn}
\end{equation}
In the non dimensionalised variables we have $\Omega(r)=r^{-s}$, and
$r_{0}$ represents $r_{0}/L$ since $L=1$. Considering finite
diffusivity $\kappa$, the form of the general solution of the evolution
equation(\ref{ndeqn}) for any initial field $\theta_{0}$ in the limit of large $t$, ie. $t\gg 1$, is the
following\cite{Flohr}
\begin{eqnarray}
\theta({\bf r},t)&= & \sum_{n} f_{n}( r,t)e^{in(\phi-\Omega(r)t)}\nonumber\\ 
f_{n}(r,t)&= & {} f_{n}(r,0)e^{[-\frac{1}{3}n^{2}\Omega^{{\prime}^{2}}Pe^{-1}t^{3}]} \nonumber\\
\label{soln1}
\end{eqnarray}
where $r=\mid{\bf r}\mid$  and $\phi$ is the azimuthal angle and $n$
is an integer. $\Omega^{\prime}$ is the derivative of $\Omega$
with respect to $r$. The angular Fourier coefficients $f_{n}( r,t)$
are time dependent and the initial condition is fully specified by
$f_{n}(r,0)$. We do not go in to the details of the solution of
(\ref{diffeqn3}) which can be found in \cite{Flohr}.
\section{Structure Functions of Passive Decaying Scalar in one
Vortex}\label{secsf} 
In this work we concentrate in finding the scaling properties of the
structure functions of the scalar field in a planar turbulence consisting
of coherent vortices. The two point equal time $p$-th order structure function is defined as follows
\begin{eqnarray}
{\cal S}_{p}(r,t)&=&\overline{\left\langle[\theta({\bf x+r},t)-\theta({\bf
x},t)]^{p}\right\rangle} \nonumber \\  
&=&\overline{\langle[\delta\theta({\bf r},t)]^{p}\rangle}. \nonumber\\ 
\label{stfunc}
\end{eqnarray}
$S_{p}$ depends only on the magnitude of the
distance between two points, when the ensemble average is
taken over an isotropic and homogeneous distribution of the scalar
field. The overbar denotes ensemble averaging and the brackets imply
space averaging ($\langle\ldots\rangle\propto\int d^{2}\bf{x}$). 

Let us first calculate $\langle[\delta\theta({\bf r},t)]^{p}\rangle$
for one 2-D vortex. We use the binomial expansion as follows
\begin{eqnarray} 
&&\langle[\theta({\bf x+r},t)-\theta({\bf x},t)]^{p}\rangle\nonumber\\&&
=\langle\theta^{p}({\bf x+r},t)\rangle+(-1)^{p}\langle\theta^{p}({\bf x},t)\rangle\nonumber\\&&
+ \sum_{q=1}^{p-1}C_{qp}(-1)^{q}\langle\theta^{p-q}({\bf x+r},t)\theta^{q}({\bf
x},t)\rangle\nonumber\\
\label{binexp} 
\end{eqnarray}
where $C_{qp}$ is the binomial coefficient of the expansion. In
order to calculate $\langle[\delta\theta({\bf r},t)]^{p}\rangle$ we first determine
\begin{equation}
\langle\theta^{q}({\bf x},t) \theta^{p-q}({\bf{x+r}},t)\rangle=B_{pq}({\bf
r},t)\
\label{stfunc1}
\end{equation}
which is the $q$-th term in the binomial expansion of
$\langle(\delta\theta({\bf r}))^{p}\rangle$ as shown in (\ref{binexp}).
Now we write the above as
\begin{equation}
\frac {1}{L_{A}^{2}}\int d^{2}{\bf x}\theta^{q}({\bf x},t) \theta^{p-q}({\bf x+
r},t)=B_{pq}({\bf r},t)\  
\label{Avg1}
\end{equation}
where $L_{A}$ is a large scale over which the spatial average may be
calculated. The Fourier transform of equation (\ref{Avg1})
is given by
\begin{equation}
\hat{F}_{pq}({\bf k},t)=\frac{1}{2\pi}\int
 e^{-i{\bf k}\cdot{\bf r}} B_{pq}({\bf r},t) d^{2}{\bf r}.
\label{Ft1}
\end{equation}
Substituting (\ref{Avg1}) in (\ref{Ft1}) and after some standard manipulations we get
\begin{eqnarray}
\hat{F}_{pq}({\bf k},t)= &&
\frac{1}{2\pi L_{A}^{2}}\int \theta^{q}({\bf x},t) e^{i{\bf k}\cdot{\bf
x}}\ d^{2}{\bf x}\nonumber\\&&
\times\int \theta^{p-q}({\bf x'},t) e^{-i{\bf k}\cdot{\bf x'}}\ d^{2}{\bf x'}.\nonumber\\\label{Ft2}
\end{eqnarray}
Now if we integrate (\ref{Ft2}) over a circular shell in $k$ space we get
\begin{equation}
F_{pq}(k,t)= \int_{0}^{2\pi}\ dA(k)\hat{F}_{pq}({\bf k},t)
\label{Ft3}
\end{equation}
where \(dA(k)\equiv kd\phi_{k}\), \(k=|{\bf {k}}|\) and $\phi_{k}$ is the
angle of ${\bf {k}}$. $F_{pq}(k,t)$ could be called the
generalised power spectrum of the scalar field in Fourier
space. Substituting (\ref{soln1}) and (\ref{Ft2}) in (\ref{Ft3}) we
get the following
\begin{eqnarray} 
&&F_{pq}(k,t)\nonumber\\&&=
\frac{1}{(2\pi L_{A}^{2})}\int dA(k)\nonumber\\&&
\times\int d^{2}{\bf x}\ e^{i{\bf k}\cdot{\bf x}}
\left\{\sum_{n}f_{n}(x,t)e^{[in(\phi-\Omega(x)t)]}\right\}^{q}\nonumber\\&&
\times \int d^{2}{\bf x'}e^{-i{\bf k}\cdot{\bf x'}}
\left\{\sum_{m}f_{m}(x',t)e^{[im(\phi'-\Omega(x')t)]}\right\}^{p-q}\nonumber\\
\label{Ft4}
\end{eqnarray}
where $x=|\bf x|$ and $x'=|\bf x'|$. After some standard manipulations
(\ref{Ft4}) leads to
\begin{eqnarray}
&&F_{pq}(k,t)\nonumber\\&&=
\frac{1}{(2\pi L_{A}^{2})}\int kd\phi_{k}\int dxx
J_{n}(kx)2\pi(i)^{n}e^{in\phi_{k}}\nonumber\\&&
\times\sum_{n,n_{1},n_{2}\ldots n_{q-1}}f_{n_{1}}f_{n_{2}}\ldots
f_{n-n_{1}\ldots -n_{q-1}}e^{-in\Omega(x)t}\nonumber\\&&
\times \int dx'x'e^{im\phi_{k}}J^{*}_{m}(kx')2\pi(-i)^{m} \nonumber\\&&
\times\sum_{m,m_{1},m_{2}\ldots m_{p-q-1}}f_{m_{1}}f_{m_{2}}\ldots
f_{m-m_{1}\ldots -m_{p-q-1}}\nonumber\\&&
\times\ e^{-im\Omega(x')t}\nonumber\\
\label{Ft5}
\end{eqnarray}
where we have changed summation variables in accordance with the following conditions
\begin{eqnarray}
&&n=n_{1}+n_{2}+n_{3}\ldots n_{q}\nonumber\\
{\rm and}\ && m=m_{1}+m_{2}+m_{3}+\ldots m_{p-q}.\nonumber\\
\end{eqnarray}
All the $f_{n}$'s and $f_{m}$'s are functions of time $t$ and of $x$
and $x'$ respectively. $J_{n}(kx)$ is  the Bessel function which has
been substituted in place of the integral representation
\begin{equation}
\int_{0}^{2\pi} e^{in\phi}e^{-ikxcos(\phi-\phi_{k})}d\phi
=2\pi(i)^{n}e^{in\phi_{k}}J_{n}(kx).
\label{Besdef}
\end{equation}
After integrating (\ref{Ft5}) with respect to $\phi_{k}$ and summing
over $m$ we get the following relation
\begin{eqnarray}
&&F_{pq}(k,t)\nonumber\\&&=
\frac{1}{(2\pi L_{A}^{2})}\int dxkx J_{n}(kx)2\pi(i)^{n} \nonumber\\&&
\times\sum_{n,n_{1},n_{2}\ldots n_{q-1}}f_{n_{1}}f_{n_{2}}\ldots
f_{n-n_{1}\ldots -n_{q-1}}e^{-in\Omega(x)t}\nonumber\\&&
\times\int dx'x'J^{*}_{-n}(kx')2\pi(-i)^{-n}\nonumber\\&&
\times\sum_{m_{1},m_{2}\ldots m_{p-q-1}}f_{m_{1}}\,f_{m_{2}}\ldots
f_{-n-m_{1}\ldots -m_{p-q-1}}\nonumber\\&&
\times\ e^{in\Omega(x')t}.\nonumber\\\label{Ft6}
\end{eqnarray}

\narrowtext \begin{figure} \epsfxsize=8truecm \epsfbox{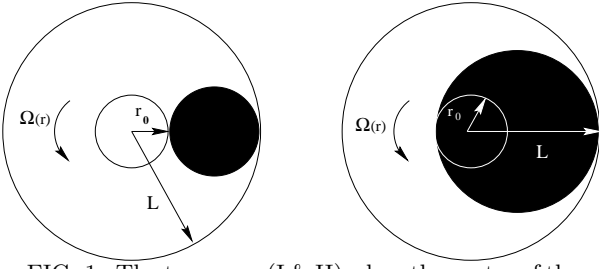}
\caption{The two cases (I \& II) when the centre of the vortex
is respectively outside a scalar patch and inside it.}
\label{f1} \end{figure}
\noindent We now have two cases to study (see figure 1). In case I the
vortex centre is outside the scalar patch and its nearest distance to
the scalar interface is $r_{0}$. In Case II the centre of the vortex
is inside the scalar patch and its nearest distance to the scalar
interface is $r_{0}$. In view of the above we can write (\ref{Ft6}) as 
\begin{eqnarray}
&&F_{pq}(k,t)\nonumber\\&&=
{\frac{1}{(2\pi L_{A}^{2})}}\left[\int_{0}^{r_{0}}+\int_{r_{0}}^{\infty}\right] dxkxJ_{n}(kx)2\pi(i)^{n}\nonumber\\&&
\times \sum_{n,n_{1},n_{2}\ldots n_{q-1}}f_{n_{1}}f_{n_{2}}\ldots
f_{n-n_{1}\ldots -n_{q-1}}e^{-in\Omega(x)t}\nonumber\\&&
\times\left[\int_{0}^{r_{0}}+\int_{r_{0}}^{\infty}\right] dx'x'J^{*}_{-n}(kx')2\pi(-i)^{-n} \nonumber\\&&
\times \sum_{m_{1},m_{2}\ldots m_{p-q-1}}f_{m_{1}}f_{m_{2}}\ldots
f_{-n-m_{1}\ldots -m_{p-q-1}}\nonumber\\&&
\times \ e^{in\Omega(x')t}.\nonumber\\
\label{Ft7}
\end{eqnarray}
In (\ref{Ft7}) we have four terms of the form as shown below
\begin{eqnarray}
&&\int_{0}^{r_{0}}dx\int_{0}^{r_{0}}dx'(\ldots)+\int_{0}^{r_{0}}dx\int_{r_{0}}^{\infty}dx'(\ldots)\nonumber\\&&
+\int_{r_{0}}^{\infty}dx\int_{0}^{r_{0}}dx'(\ldots)+\int_{r_{0}}^{\infty}dx\int_{r_{0}}^{\infty}dx'(\ldots)\nonumber\\
\label{sumint}
\end{eqnarray}
with the integrands denoted by $(\ldots)$ being the same as (\ref{Ft7})
for all the terms of the above. For case I, terms containing
\(\int_{0}^{r_{0}}(\ldots)\) are zero since \(f_{n}=0\) in the regime \(0<x<r_{0}\) for all n,
even for $n=0$, since there is no scalar patch in the region
\(0<x<r_{0}\). Therefore contributions only come from the term
\[\int_{r_{0}}^{\infty}dxx\int_{r_{0}}^{\infty}dx'x'(\ldots).\]
For case II we can legitimately replace $\theta$ by $\theta-f_{0}$ and
get the same result as in case I. Hence the only contributing  term is
the following  
\begin{eqnarray}
&&F_{pq}(k,t) \nonumber\\&&= 
\frac{1}{(2\pi L_{A}^{2})}\int_{r_{0}}^{\infty} dxkxJ_{n}(kx)2\pi(i)^{n}\nonumber\\&&
\times \sum_{n,n_{1},n_{2}\ldots n_{q-1}}f_{n_{1}}f_{n_{2}}\ldots
f_{n-n_{1}\ldots -n_{q-1}}e^{-in\Omega(x)t}\nonumber\\&&
\times \int_{r_{0}}^{\infty} dx'x'J^{*}_{-n}(kx')2\pi(-i)^{-n}\nonumber\\&&
\times\sum_{m_{1},m_{2}\ldots
m_{p-q-1}}f_{m_{1}}f_{m_{2}}\ldots f_{-n-m_{1}\ldots -m_{p-q-1}}\nonumber\\&&
\times \ e^{in\Omega(x')t}.
\label{Ft8}
\end{eqnarray}
It is because $f_{0}$ decays with a time scale which is much larger
than the decay time of the non-zero harmonics\cite{Gilbert} that
$f_{0}$ is considered to be a constant and therfore subtracted away
from the $\theta$ in case II. The same reasoning can be applied for
case I. Hence all the $n_{i}$'s and $m_{i}$'s are non-zero in the
above equation and in the rest of the paper .  

To take into account the fact that diffusion gradually wipes out the spiral
structure of the scalar field near the vortex centre (a fact not taken
in to account in \cite{Gilbert} where the spiral structure is assumed
to exist wholly intact until finally destroyed by viscosity), we follow Flohr
and Vassilicos\cite{Flohr} and
define a critical radius $\rho$ which gives a measure of this diffused
region. This critical radius is defined in the limit
$Pe\rightarrow\infty$ for times $t\ll Pe^{1/3}$ which are such
that \[f_{n}(r,t)=f_{n}(r,0)\ for\ r\ll\rho\]
but \[ |f_{n}(r,t)|\ll |f_{n}(r,0)|\ for\ r\gg\rho\ see
(\ref{soln1}).\]
Hence, we set
\(\frac{1}{3}n^{2}\Omega'^{2}(\rho)Pe^{-1}t^{3}=1\) which implies
\[\rho(t)=\left[\frac{1}{3}n^{2}s^{2}Pe^{-1}t^{3}\right]^{\frac{1}{2(s+1)}}.\]
This critical radius is time-dependent and grows with time.
It can be thought of as a diffusive length scale over which the
harmonics in $n$ have diffused and the spiral
structure has been smeared out.

In view of the above, the integrals in (\ref{Ft8}) can be further divided as
\[\int_{r_{0}}^{\infty}dx =\int_{r_{0}}^{\rho}dx+\int_{\rho}^{\infty}dx.\]
The only significantly non-zero contribution comes
from the range \(\rho<x<1\) in the integrals and similarly for
$x'$. Hence we get the following
\begin{eqnarray}
&&F_{pq}(k,t)\nonumber\\&&= 
\frac{1}{(2\pi L_{A}^{2})}\int_{\rho}^{\infty} dxkxJ_{n}(kx)2\pi(i)^{n}\nonumber\\&&
\times
\sum_{n,n_{1},n_{2}\ldots n_{q-1}}f_{n_{1}}f_{n_{2}}\ldots f_{n-n_{1}\ldots
-n_{q-1}}e^{-in\Omega(x)t}\nonumber\\&&
\times \int_{\rho}^{\infty} dx'x'J^{*}_{-n}(kx')2\pi(-i)^{-n}\nonumber\\&&
\times\sum_{m_{1},m_{2}\ldots m_{p-q-1}}f_{m_{1}}f_{m_{2}}\ldots
f_{-n-m_{1}\ldots -m_{p-q-1}}\nonumber\\&&
\times \ e^{in\Omega(x')t}
\label{Ft14}
\end{eqnarray}
Now for large $kx$, ie. $kx\gg 1$, we can use the asymptotic expansion for the Bessel function
\begin{equation}
J_{n}(kx)\sim \left(\frac{1}{2\pi kx}\right)^{\frac{1}{2}}\left[(-i)^{n+\frac{1}{2}}e^{ikx}+(i)^{n+\frac{1}{2}}e^{-ikx}\right].        
\label{Bes}
\end{equation}
This is appropriate for our analysis if the Fourier modes are to resolve
at the very least the distance $r_{0}$ from the centre of the vortex
to the scalar patch interface i.e \(1<kr_{0}\).
After substituting (\ref{Bes}) in (\ref{Ft14}) we use the method of
stationary phase to evaluate the integrals where the phase is given by
\begin{equation}
\Phi=kx-n\Omega(x)t.\
\label{phase}
\end{equation}
The approximation for a general integral of this type is known to be
\begin{eqnarray}
I(x)&= & \int_{a}^{b}f(t)exp[ix \Psi(t)]dt \nonumber\\
    &\sim & {}\sqrt{\frac{\pi}{2x\mid\Psi''(t^{\ast})\mid}}f(t^{\ast})exp[ix\Psi(t^\ast)\pm\pi/4]\nonumber\\
\label{stphse}
\end{eqnarray}
where $t^{*}$ is the t where the derivative of the phase is zero. The
condition of stationary phase gives
\begin{equation}
0=\Phi'=-k-n\Omega'(x_{n})t
\label{phase1}
\end{equation}
which picks out points $x_{n}$ where the contribution to the integral
is maximum. Finally what we get is
\begin{eqnarray}
\lefteqn{F_{pq}(k,t)\sim} \nonumber\\
& & {}\sum_{n,n_{1}\ldots n_{q-1}}\frac{2\pi k}{n|\Omega''(x_{n})|t}\nonumber\\
& & {} \times\left(\frac{x_{n}}{2\pi k}\right)f_{n_{1}}f_{n_{2}}\ldots f_{n-n_{1}-n_{2}-\ldots -n_{q-1}} \nonumber\\
& & {} \times\sum_{m_{1},m_{2}\ldots m_{p-q-1}}
f_{m_{1}}f_{m_{2}}\ldots f_{-n-m_{1}-m_{2}-\ldots -m_{p-q-1}} \nonumber\\
\label{Ft9}
\end{eqnarray}
where \(f_{n}=f_{n}(x_{n},0)\) and similarly for $f_{m}$.
Now from the condition of stationary phase (\ref{phase1}) we can find
\begin{equation}
x_{n}=\left(\frac{snt}{k}\right)^{\frac{1}{s+1}}
\label{phase2}
\end{equation}
where we have used \(\Omega(r)= r^{-s}\). The stationary phase
contributes only when \(\rho<x_{n}<1\) because the spiral structure
only exists in that range of distances from the centre of the vortex.
The relation between the fractal co-dimension (Kolmogorov capacity)
$D$ of the scalar spiral and the power law of the decay of the azimuthal
velocity of the vortex is\cite{Flohr},\cite{Vas91}
\begin{equation}
D=\frac{s}{s+1}. \label{fract}
\end{equation}
This $D$ is such that \(1/2<D<2/3\) because $1<s<2$ and gives a measure of the space-filling property of the spiral. Hence after doing the summations in (\ref{Ft9}) we can show that the power spectrum $F_{pq}(k,t)$ scales like
\begin{eqnarray}
&&F_{pq}(k,t)\nonumber\\&&
\sim k^{-(3-2D)} t^{2(1-D)}[\rm{const+\ higher\ order\ terms}]\nonumber\\
\label{Ft10}         
\end{eqnarray}
in the limit $Pe\rightarrow\infty$ and in the range \( t<k<\sqrt{\frac{Pe}{t}}\) for times
\(1\ll t\ll Pe^{1/3}\) which is the range of times for which the
scalar patch has a well-defined spiral structure in the
range of wavenumbers \(t<k<\sqrt{{\frac{Pe}{t}}}\)(obtained from $\rho<x_{n}<1$). The higher order terms
are functions of $k/t$, and decay faster than $(k/t)^{-1}$ in the
range \(t<k<\sqrt{\frac{Pe}{t}}\), and can therefore be neglected.

We notice that as time runs forward the spiral range
\(\left(t<k<\sqrt{\frac{Pe}{t}} \right)\) shifts to higher values of
$k$ which is solely due to the vortex
continuously wrapping the scalar field in to finner and finner spirals
thus generating scales which have higher wavenumbers. This range
also shrinks as it shifts to higher values of $k$ because of the
action of diffusion. In the next section we generalise our results
to the case of multiple spirals generated by a dilute collection of
noninteracting vortices which may be representative of a turbulent
velocity field with coherent vortical structure, perhaps obtained in the
experiments of Jullien et al.\cite{JCT99}.

\section{Structure functions of passive decaying scalar in a flow consisting of
many identical non-interacting vortices}\label{secmulsp}  
All the analysis in this section is carried out in dimensionalised
variables so we invert the transformations of section
\ref{sec1vor}. Let us consider many non interacting vortices
randomly distributed over 2-D space and sufficiently far apart so that
we can safely describe the velocity field in terms of compact vortical
structures characterised by 
\[\Omega(x)=\Omega_{0}\left({\frac{x}{L}}\right)^{-s}\ {\rm if}\ {\frac{x}{L}}\leq 1\]
\[\Omega(x)=0\ {\rm if}\ {\frac{x}{L}}>1\]
where the $x$'s are measured from the centre of each vortex at
${\bf{x}}_{m}$ for all $m$ and 
\begin{equation}
min|{\bf{ x}}_{m}-{\bf{x}}_{n}|\gg L\ ;{\rm{for\ all\ m\ and\ n.}}
\label{discond}
\end{equation}
For the calculation of the generalised power spectrum and structure
functions we need only to consider the scalar patches within a
distance $L$ of each vortex because these scalar patches acquire a
spiral structure and thereby dominate the scaling of the structure
functions. The scalar field at distances larger than $L$ from all vortices
contributes an ${\cal O}(r/L)$ term to the structure function for
\(r\ll L\) because the interfacial structure of the scalar field far
from the vortices remain regular. As we show in this section, this
term is negligible in the \(r/L\ll 1\) limit compared to the
$r$-dependence of the structure functions caused by the scalar spiral
structures around the vortices. It is therefore sufficient to consider
that $\theta({\bf x},t)$ consists only of the local scalar fields
$\theta_{m}({\bf x-x}_{m},t)$ in the vicinity of vortices labelled $m$
and write
\begin{eqnarray}
\theta({\bf x},t)=\sum_{m}\theta_{m}({\bf{ x-x}}_{m},t).
\label{mulvort1}
\end{eqnarray} 
\noindent Every scalar spiral in the right hand side of
(\ref{mulvort1}) is localised within a distance $L$ of ${\bf{x}}_{m}$ and the
condition (\ref{discond}) ensures that they do not overlap
each other. 

Now we can generalise (\ref{Ft2}) to include the effect of many
non-interacting vortices with non-overlapping scalar spirals as shown
below
\begin{eqnarray}
\hat{F}_{pq}({\bf k},t)&=&
\frac{1}{2\pi L_{A}^{2}}\sum_{m}\int\theta_{m}^{q}({\bf
x},t) e^{i{\bf k}\cdot{\bf x}}d^{2}{\bf x}\nonumber\\
&&\times \int \theta_{m}^{p-q}({\bf x'},t) e^{-i{\bf k}\cdot{\bf x'}}d^{2}{\bf x'}.
\label{mulft1}
\end{eqnarray}
Since the integrals are independent of the $m$'s
we can take them out of the sum. Hence  we get the same result as
in (\ref{Ft10}) multiplied by the number of vortices per unit area,
that is
\begin{equation}
F_{pq}(k,t)\sim(kL)^{-(3-2D)}(\Omega_{0}t)^{2(1-D)}\sum_{m}\frac{1}{2 \pi L_{A}^{2}}.
\label{mulft2}
\end{equation}
This asymptotic relation is valid when
\(\Omega_{0}t<kL<\sqrt{\frac{Pe}{\Omega_{0}t}}\) which is found from
the condition \(\rho<x_{n}<l\) in dimensionalised form and
\(1\ll\Omega_{0}t\ll Pe^{1/3}\) in the limit $Pe\rightarrow\infty$.

Assuming the distribution of the scalar spirals over the two
dimensional space to be homogeneous and isotropic, the power spectrum
\(F_{pq}(k,t)=2\pi k\overline{\hat{F}_{pq}({\bf k},t)}\) where the bar
implies ensemble averaging over the distribution of many spirals
(because vortices are non-interacting and spirals are therefore
statistically independent from each other, it does make sense for the
average over space already included in the definition of
$\hat{F}_{pq}({\bf k},t)$ to be taken over an idealised space where
there is only one spiral, and for the ensemble average to be taken over
the distribution of many spirals).  From (\ref{binexp}) we can show
that for a homogenous distribution of scalar spirals the odd order
structure functions vanish. Only the even order structure functions do
not vanish, that is for $p=even$.   Hence from(\ref{stfunc})  
\begin{eqnarray}
{\cal S}_{p}(r,t)&= & \overline{\langle[\theta({\bf x+r},t)-\theta({\bf x},t)]^{p}\rangle}\nonumber\\
& =& {}\overline{\langle[\delta\theta({\bf r},t)]^{p}\rangle} \nonumber\\
& \sim& {}\int\left(2-\sum_{q=1}^{p-1}C_{qp} e^{i{\bf k}\cdot{\bf r}}\right){\overline{\hat{F}_{pq}({\bf{k}},t)}}kdkd\phi \nonumber\\
& \sim& {}\int\left(1-e^{i{\bf k}\cdot{\bf r}}\right)F_{pq}(k,t)dkd\phi \nonumber\\
& \sim& {}\int(1-J_{0}(kr))F_{pq}(k,t)dk\nonumber\\
\label{stfunc2}
\end{eqnarray}
where $J_{0}(kr)$ is same as (\ref{Besdef}) with \(n=0\).
After substituting (\ref{mulft2}) in (\ref{stfunc2}) and integrating
we find 
\begin{eqnarray}
&&{\cal{S}}_{p}(r,t)\sim \left({\frac{r}{L}}\right)^{2(1-D)}(\Omega_{0}t)^{2(1-D)}\nonumber\\&&
{\rm{in\ the\ ranges}}\
\frac{1}{\Omega_{0}t}>\frac{r}{L}>\sqrt{\frac{\Omega_{0}t}{Pe}}\nonumber\\&&
{\rm{ and}}\ 1\ll\Omega_{0}t\ll Pe^{1/3}.
\label{stfunc3}
\end{eqnarray}


\section{Structure functions of statistically stationary passive
scalar}\label{avltm}
To achieve a statistically stationary passive scalar field we may
imagine that, as scalar patches take spiral shapes and decay, more
scalar patches are introduced in to the flow as may well happen in an
experimental setup in the laboratory. This procedure soon leads to a
situation where many scalar spirals coexist in the flow all in
different stages of their evolution. Assuming the rate of regular
injection of the scalar blobs to balance exactly the rate of scalar
dissipation, we can expect to have a statistically stationary scalar
field. In this case, the averaging over many spirals in different
stages of their evolution (which is involved in the calculation of the
generalised power spectra and structure functions) may be assumed, in
the spirit of Lundgren\cite{Lund82}, to be equivalent to averaging
over the life-time of a single spiral. Hence, to obtain the
generalised power spectra of the statistically stationary scalar we
average the previous section's results over time in the range
\(1<\Omega_{0}t<Pe^{1/3}\), which represents the life time of the
spirals. The spiral structure lies in the range \(\rho<x_{n}<L\) which
implies \(\Omega_{0}t<kL<\sqrt{\frac{Pe}{\Omega_{0}t}}\). This spiral
range of wavenumbers together with the time range of the spiral gives
the range of $\Omega_{0}t$ over which we can average for a given value of
$kL$. This leads to a time averaged $F_{pq}(k,t)$ which takes the form
\begin{eqnarray}
F_{pq}(k)&\sim &(kL)^{-1}\nonumber\\
{\rm{ where}}&& 1< kL< Pe^{1/3},
\label{Ft13}
\end{eqnarray}
\begin{eqnarray}
&F_{pq}(k)\sim (kL)^{-(7-6D)}Pe^{2(1-D)} &\nonumber\\
{\rm{ where}}& Pe^{1/3}< kL< Pe^{1/2}.
\label{Ft15}
\end{eqnarray}
(\ref{Ft13}) and (\ref{Ft15}) are the result of averaging
(\ref{mulft2}) over the time ranges \(1<\Omega_{0}t<kL\) and
\(1<\Omega_{0}t<\frac{Pe}{(kL)^{2}}\) respectively. These time ranges are
determined by the respective wavenumber ranges in (\ref{Ft13})
and (\ref{Ft15}). Finally (\ref{Ft13}) and (\ref{Ft15}) give the
following structure functions 
\begin{eqnarray}
&&{\cal S}_{p}(r)\sim constant+\ln\left({\frac{r}{LPe^{-1/3}}}\right) \nonumber\\&&
{\rm{ where}}\ LPe^{-1/3}<r<L ,
\label{stfunc4}
\end{eqnarray}
\begin{eqnarray}
{\cal S}_{p}(r)&\sim & \left({\frac{r}{LPe^{-1/3}}}\right)^{6(1-D)}\nonumber\\
{\rm{ where}}\ &&LPe^{-1/2}<r<LPe^{-1/3}.
\label{stfunc5}
\end{eqnarray}
In (\ref{stfunc4}) and (\ref{stfunc5}) ${\cal{S}}_{p}(r)$ is a time average of
${\cal{S}}_{p}(r,t)$ in (\ref{stfunc3}). Note that $\zeta_{p}=0$ with a
logarithmic correction in the range \(LPe^{-1/3}<r<L\) and
that \(\zeta_{p}=6(1-D)\in]2,3[\) in the dissipative range
\(LPe^{-1/2}<r<LPe^{-1/3}\). 

The structure functions ${\cal{S}}_{p}(r)$ are not time-dependent and may be
interpreted as characterising a scalar field in a statistically steady
state achieved with an external large-scale source of
scalar (scalar forcing). This scalar forcing may consist of regularly
placing in the flow scalar blobs with large-scale smooth interfacial
structure. In the spirit of Birkhoff's Ergodic theorem \cite{Frisch}
we should expect Lundgren's time average assumption to be relevant for
the calculation of structure functions when the scalar field is
statistically stationary.

\section{Phenomenology}\label{pheno}
In this section we extract the phenomenology underlying the calculations
and results of the previous sections and show that the essential
ingredient of this phenomenology are the locality of scalar
inter-scale transfer (\ref{spctr1}) and the Lundgren time-averaging
operation. Indeed, as we show in this section, (\ref{Ft13}) and
(\ref{Ft15}) can be retrieved by a simple phenomenological argument
based on these two ingredients.

Let us return to the time-dependent wind-up of scalar spirals.
As time proceeds, ie. as \(\Omega_{0}t\rightarrow\Omega_{0}(t+\delta t)\),
then \(kL\rightarrow kL+L\delta k\) because of the differential
rotation(which amounts to local shear) in every steady vortex and the
entire scalar spectrum is shifted towards higher wavenumbers with
time (see Figure 2). That is to say that the shearing advection to which the
scalar patches are subjected in every steady vortex is such that the
generalised power spectra obey
\begin{eqnarray}
&&F_{pq}(kL+\delta kL,\Omega_{0}t+\Omega_{0}\delta t)d(kL+\delta
kL)\nonumber\\&&
=F_{pq}(kL,\Omega_{0}t)d(kL)
\label{spctr1}
\end{eqnarray}
\narrowtext \begin{figure} \epsfxsize=8truecm \epsfbox{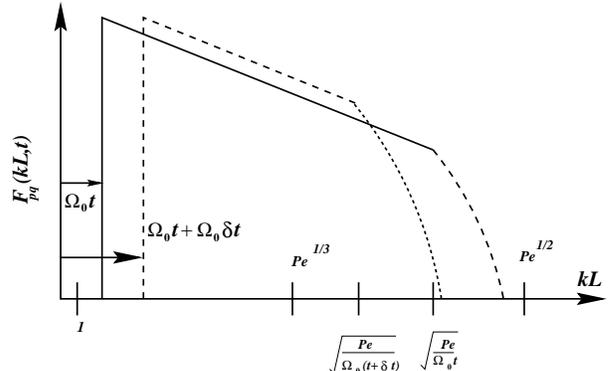}
\caption{The time evolution of the scalar power spectrum
$F_{pq}(kL,t)$  in a log-log plot.}
\label{f2} 
\end{figure}
The amount of scalar variance in the wavenumber band $d(kL)$ around
wavenumber $kL$ is simply transported to wavenumber band $d(kL+\delta
kL)$ around wavenumber $kL+\delta kL$ after an incremental time
duration $\Omega_{0}\delta t$ (see Figure 2). As shown in\cite{99AV},
the distance $l$ between consecutive  coils of the scalar spiral in every
vortex at a distance $r$ from the centre scales as
\begin{equation}
\l\sim \frac{L}{\Omega_{0}t}\left(\frac{r}{L}\right)^{1+1/s}.
\label{ltscal}
\end{equation}
Letting time vary by a small amount $\delta t$, the distance between
two coils changes by
\[\frac{\delta l}{l}\simeq -\frac{\delta t}{t}\]
the minus sign indicating that $l$ is decreasing. Identifying $k$
with $\frac{2\pi}{l}$ for the purpose of equation
(\ref{spctr1}) so that \(\frac{\delta k}{k}= -\frac{\delta
l}{l}\), it follows that (\ref{spctr1}) becomes 
\begin{eqnarray}
&&F_{pq}\left[kL\left(1+\frac{\delta t}{t}\right),\Omega_{0}t\left(1+\frac{\delta
t}{t}\right)\right]dkL\left(1+\frac{\delta t}{t}\right)\nonumber\\&&
=F_{pq}\left(kL,\Omega_{0}t\right)d(kL).
\label{spctr2}
\end{eqnarray} 
The solution of this equation is
\begin{equation}
F_{pq}(kL,\Omega_{0}t)=L(kL)^{-1}{\cal F}_{pq}\left(\frac{\Omega_{0}t}{kL}\right)
\label{spctr3}
\end{equation}
where ${\cal{F}}_{pq}$ are arbitrary dimensionless functions. As
indicated in figure 2 this form of the generalised spectra is valid in
the limit \(Pe\rightarrow\infty\) and in the wavenumber range
\(\Omega_{0}t\ll kL\ll\sqrt{\frac{Pe}{\Omega_{0}t}}\) and time range
\(1\ll\Omega_{0}t\ll Pe^{1/3}\). Note that \(1\ll\Omega_{0}t\ll
Pe^{1/3}\ll\sqrt{\frac{Pe}{\Omega_{0}t}}\ll Pe^{1/2}\). The inverse of
$\Omega_{0}t$ represents the decaying outer length-scale of the spiral
range and the inverse of $\sqrt{\frac{Pe}{\Omega_{0}t}}$ represents
the growing micro-scale of diffusive attrition. A wavenumber in the
range \(1\ll kL\ll Pe^{1/3}\) during the time-period
\(1\ll\Omega_{0}t\ll Pe^{1/3}\) does not have the time to be affected
by diffusive attrition and only receives scalar variance activity from
lower wavenumbers until $\Omega_{0}t=kL$. We therefore refer to \(1\ll
kL\ll Pe^{1/3}\) as the advective wavenumber range. The time averaged
generalised power spectra in this range are given by
\begin{equation}
F_{pq}(kL)=\frac{1}{kL-1}\int_{1}^{kL}d(\Omega_{0}t)L(kL)^{-1}{{\cal{F}}_{pq}}\left(\frac{\Omega_{0}t}{kL}\right)
\label{power1}
\end{equation}
and ${\cal{F}}_{pq}$ must be increasing functions of $\frac{\Omega_{0}t}{kL}$
because the differential rotation's shearing process causes the power
spectra to shift from small to large wavenumbers (see figure 2). Hence 
we retrieve (\ref{Ft13}), i.e.
\[F_{pq}(kL)\sim (kL)^{-1}\]
in the advective range \(1\ll kL\ll Pe^{1/3}\) which is well defined
in the limit $Pe\rightarrow\infty$.

In the advective-diffusive range  \(Pe^{1/3}\ll kL\ll Pe^{1/2}\) a
wavenumber $kL$ experiences the advection process from $\Omega_{0}t=1$
until $kL=\sqrt{\frac{Pe}{\Omega_{0}t}}$ when molecular diffusion sets
in. Hence the time averaged generalised power spectra are given by 
\begin{equation}
F_{pq}(kL)=\frac{1}{\frac{Pe}{kL^{2}}-1}\int_{1}^{\frac{Pe}{kL^{2}}}d(\Omega_{0}t)L(kL)^{-1}{{\cal{F}}_{pq}}\left(\frac{\Omega_{0}t}{kL}\right)
\label{power2}
\end{equation}
in the advective diffusive range and using \(
{\cal{F}}_{pq}\left(\frac{\Omega_{0}t}{kL}\right)\sim\left(\frac{\Omega_{0}t}{kL}\right)^{2(1-D)}\)
 (see (\ref{mulft2})) we retrieve (\ref{Ft15}), i.e.
\[F_{pq}(kL)\sim Pe^{2(1-D)}(kL)^{-7+6D}\]
in the advective-diffusive range \(Pe^{1/3}\ll kL\ll Pe^{1/2}\) which
is well defined in the limit $Pe\rightarrow\infty$. Note that the
diffusive micro-length-scale $LPe^{-1/2}$ is the Taylor microscale of
the scalar field (first introduced by Corrsin  
in 1951). Note also that $7-6D\in]3,4[$ and that the
experimental results of \cite{JCT99} seem to show a steeper power-law
wavenumber spectrum at wavenumbers larger than where the $k^{-1}$ spectrum is observed.

\section{Conclusions and Discussion}\label{disc}

In a  two-dimensional isotropic and homogeneous collection of
non-interacting compact and time-independent singular vortices with a
large-wavenumber energy spectrum $E(k)\sim k^{-\alpha}$ with
$1<\alpha\leq 3$, the structure functions of an advected and freely
decaying scalar field have the following scaling
behaviour in the limit where $Pe\rightarrow\infty$
\begin{equation}
{\cal{S}}_{p}(r,t)\sim
\left({\frac{r}{L}}\Omega_{0}t\right)^{2(1-D)}
\label{stfunc7}
\end{equation}
where \(\sqrt{\frac{\Omega_{0}t}{Pe}}\ll\frac{r}{L}\ll{\frac{1}{\Omega_{0}t}}\)
and \(1\ll\Omega_{0}t\ll Pe^{1/3}\), and the fractal co-dimension $D$
of the scalar interfaces is such that $1/2\leq D<2/3$.

By applying the Lundgren time-average assumption we obtain predictions
for the structure functions of a statistically stationary scalar
field in the same 2-D velocity field and the same limit $Pe\rightarrow\infty$
\begin{equation}
{\cal S}_{p}(r)\sim constant+\ln\left({\frac{r}{LPe^{-1/3}}}\right)
\label{stfunc8}
\end{equation}
in the range \( LPe^{-1/3}\ll r\ll L\) and
\begin{equation}
{\cal S}_{p}(r)\sim\left({\frac{r}{LPe^{-1/3}}}\right)^{6(1-D)}
\label{stfunc9}
\end{equation}
in the range \(LPe^{-1/2}\ll r\ll LPe^{-1/3}\). The logarithmic term in (\ref{stfunc8}) corresponds to $k^{-1}$
generalised power spectra. It may be worth mentioning that the 2-D
velocity fields of Holzer and Siggia\cite{HolSig} where they observe a
well-defined $k^{-1}$ scalar power spectrum are replete with spiral
scalar structures.

Predictions of $k^{-1}$ scalar power spectra in the limit
$Pe\rightarrow\infty$ have been made for scalar fields in smooth
(ie. at least everywhere continuous and differentiable ) homogeneous
and isotropic random velocity field  with arbitrary dimensionality and
time dependence by Chertkov et. al.\cite{Cher} who generalised and
unified the results of Batchelor\cite{Bat59} and Kraichnan
\cite{Kr68}. Experimental investigations of the high P\'eclet number
$k^{-1}$ scalar power spectrum have been inconclusive in 3-D turbulent
flows even though Prasad and Sreenivasan have claimed such a spectrum in a 3-D wake \cite{MD96}. However $k^{-1}$ scalar power spectra have been
observed at high P\'eclet numbers in numerical simulations of scalar
fields in 2-D and 3-D chaotic flows \cite{CVul90,AnFen95,An96} and in
2-D velocity fields obeying the stochastically forced Euler equation
restricted to a narrow band of small (integral scale) wavenumbers
\cite{HolSig}. More recently, $k^{-1}$ scalar power spectra have been
observed at $Pe=10^{7}$ in 2-D or quasi 2-D statistically stationary
turbulent flows in the same range where the velocity field's energy
spectrum is $k^{-3}$ by Jullien et al.\cite{JCT99} who have also
observed logarithmic scalings in that range for all order structure
functions (similarly to (\ref{stfunc8}), but without the ability to
establish the $LPe^{-1/3}$ scaling factor and range). The theory of
Chertkov et al.\cite{Cher} does not apply to this experiment because
homogeneous and isotropic random velocity fields which are everywhere
continuous and differentiable have energy spectra steeper than
$k^{-4}$ (see Appendix). The present paper's theory, however, applies
when the energy spectrum  scales as $k^{-\alpha}$ with $1<\alpha\leq 3$
but is limited to time-independent velocity fields. Nevertheless the
phenomenology developed in section \ref{pheno} also holds for
time-dependent velocity fields and we now apply and generalise it to
spatially smooth chaotic flows (and also, by the way, to frozen
straining velocity field structures).

The starting point of our phenomenology is the locality of transfer
(\ref{spctr1}). Pedrizzetti and Vassilicos \cite{PV2000} have shown that
inter-scale transfer in 2-D compact vortices is indeed local at a
given scale when velocity gradients do not vary much in physical space
over that scale. This is the case in the 2-D axisymmetric vortices
considered in this paper but also in spatially smooth velocity fields.
In a spatially smooth chaotic flow the distance $l$ between successive
folds of the scalar interface decays exponentially as determined by
the largest positive Lyapunov exponent $\lambda$, ie. \(l(t)\sim
e^{-\lambda t}\) which implies \({\frac{\delta
l}{l}}\simeq-\lambda\delta t\). Applying the locality of transfer
property we get
\begin{eqnarray}
&&F_{pq}(k(1+\lambda\delta t),t+\delta t)dk(1+\lambda\delta t)\nonumber\\&&
=F_{pq}(k,t)dk
\label{spctr4}
\end{eqnarray}
the solution of which is
\begin{equation}
F_{pq}(k,t)= k^{-1}{\cal F}_{pq}\left({\frac{e^{\lambda t}}{k}}\right).
\label{spctr5}
\end{equation}
This form of the generalised spectra is valid for
$Pe\rightarrow\infty$ and as long as \(1<e^{\lambda t}<k\), so that
applying Lundgren's time-average operation from $t=0$ to
\(t=\lambda^{-1}\ln k\) gives
\begin{eqnarray}
F_{pq}(k)
&=& k^{-1}{\frac{\lambda}{\ln k}}\int_{0}^{{\frac{\ln k}{\lambda}
}}{\cal{F}}_{pq}\left(\frac{e^{\lambda t}}{k}\right)dt\nonumber\\
&\sim & [k\ln k]^{-1}
\label{spectr6}
\end{eqnarray}
because ${\cal{F}}_{pq}$ is an increasing function of \(
\frac{e^{\lambda t}}{k}\). Power spectra of scalar fields in
spatially smooth chaotic flows are believed to scale as $k^{-1}$ in
the limit $Pe\rightarrow\infty$ \cite{CVul90,AnFen95,An96} but our
theory predicts $(k\ln k)^{-1}$. This is a small correction to the
spectrum but an exponentially large  correction to the scalar variance in the
limit $Pe\rightarrow\infty$.

\section{Acknowledgments}
MAIK and JCV acknowledge support from EPSRC grant GR/K50320 and from EC
TMR Research network on intermittency in turbulent systems. MAIK also
wishes to thank the Cambridge Commonwealth Trust, Wolfson College,
Cambridge and DAMTP for financial support while this work was being
completed. JCV acknowledges support from the Royal Society.

\section{Appendix} 
For a statistically homogeneous velocity field with velocity
components $u_{i}({\bf x})$ we can define a correlation function 
\[ R_{ij}({\bf r})=\overline{\langle u_{i}({\bf x})u_{j}({\bf x}+{\bf
r})\rangle}\]
and its Fourier transform
\[\Phi_{ij}({\bf k})={\frac{1}{(2\pi)^{2}}}\int d{\bf r}R_{ij}({\bf r})e^{-i{\bf k}\cdot{\bf r}}.\]
Incompressibility ($\nabla\cdot{\bf u}=0$) and statistical isotropy of
a two-component or 2-D velocity field $u_{i}({\bf{x}})$, $i=1,2$,
imply 
\[\Phi_{ij}({\bf
k})=\left(\delta_{ij}-\frac{k_{i}k_{j}}{k^{2}}\right)\frac{E(k)}{\pi
k}\]
where the average kinetic energy per unit mass of the velocity field is
\(E=\int_{0}^{\infty} dk E(k), k\equiv|{\bf k}|\); $E(k)$ is the
energy spectrum of the velocity field.

One dimensional energy spectra are defined as follows
\[\phi_{ij}(k_{1})=\frac{1}{2\pi}\int_{-\infty}^{\infty} R_{ij}(r_{1},0)e^{-ik_{1}r_{1}}dr_{1}\]
which, because of isotropy, are completely characterised by a single
component, say $ \phi_{11}(k_{1})$. 

From the above and a few standard manipulations one can get
\[\int_{1}^{\infty}dx
\frac{\sqrt{x^{2}-1}}{\pi x^{2}}\frac{E(xk_{1})}{\phi_{11}(k_{1})}=\frac{1}{2}\]
and in a range of wavenumbers where both $E(k)$ and $\phi_{11}(k_{1})$
are monotonically decreasing functions of $k$ and $k_{1}$
respectively, it follows that \(E(k)\sim k^{-p}\) if and only if
\(\phi_{11}(k_{1})\sim k_{1}^{-p}\).

The pivotal assumption in \cite{Cher} is that the velocity field
$u_{i}({\bf x})$ is Taylor-expandable up to at least first derivative
terms everywhere in physical space. In particular, this means that
$u_{1}(x_{1},0)$ is differentiable with respect to $x_{1}$ everywhere
on the $x_1$ axis.  If the first derivative of $u_{1}(x_{1},0)$ with
respect to $x_{1}$ is also continuous everywhere along the $x_1$ axis
then the Fourier transform $\hat{u}_{1}(k_{1})$ of $u_{1}(x_{1},0)$
must decay faster than $k_{1}^{-2}$ \cite{CourHilbrt53}. If, however,
the first derivative of $u_{1}(x_{1},0)$ is not everywhere continuous,
then it is discontinuous either on a set of well-separated points or
on a more pathological set of points which accumulate (and are
therefore not well-separated) in a fractal-like or in a spiral-like
manner \cite{Vas91,CourHilbrt53}. In the case where discontinuities in
the derivative field of $u_{1}(x_{1},0)$ are well-separated,
$\hat{u}_{1}(k_{1})$ decays as $k_{1}^{-2}$ because the Fourier
transform of well-separated discontinuities between which the field is
continuous decays as $k_{1}^{-1}$ and the Fourier transform of the
derivative of $u_{1}(x_{1},0)$ is equal to $ik_{1}
\hat{u}_{1}(k_{1})$. In the other case where discontinuities are not
well-separated and accumulate, the decay of $\hat{u}_{1}(k_{1})$ can
be anywhere between $k_{1}^{-1}$ and $k_{1}^{-2}$
\cite{CourHilbrt53,BelchVas97}, but in this case $u_{1}(x_{1},0)$
cannot be considered to be differentiable at those points where
discontinuities of its derivative accumulate.\\

In conclusion, the differentiability of the velocity field everywhere
in physical space implies that $\hat{u}_{1}(k_{1})$
must decay at least as $k_{1}^{-2}$ and therefore
$\phi_{11}(k_{1})\sim \vert \hat{u}_{1}(k_{1})\vert^{2}\sim
{\cal O}(k_{1}^{-4})$ which in turn implies $E(k)\sim {\cal O}(k^{-4})$.\\

The condition \(E(k)\sim{\cal{O}}(k^{-3})\) stated in the conclusion
of \cite{Cher} guarantees that the strain field is large-scale but not
that the velocity field is differentiable. The spectral condition
required to use the pivotal assumption of differentiability in
\cite{Cher} should in fact be \(E(k)\sim{\cal{O}}(k^{-4})\).


\end{multicols}
\end{document}